\documentclass{aa}

\usepackage{amssymb}
\usepackage{amsmath}
\usepackage{orcidlink}
\usepackage{color}
\usepackage{natbib}
\bibpunct{(}{)}{;}{a}{}{,}
\usepackage{hyperref}

\bibpunct{(}{)}{;}{a}{}{,}

\usepackage{graphicx}

\usepackage[varg]{txfonts}

\begin{document}

 \title{From reflection to scattering: Polarimetric signatures of funnel-type outflows}

   \subtitle{Modeling obscured ultraluminous X-ray sources}

   \titlerunning{Obscured ultraluminous X-ray sources}

   \author{Varpu Ahlberg,
          \inst{1}\orcidlink{0009-0006-9714-5063}
          Alexandra Veledina,
          \inst{1,2}\orcidlink{0000-0002-5767-7253} 
          Eugene Churazov,
          \inst{3,4} \orcidlink{0000-0002-0322-884X}
          Ildar Khabibullin
          \inst{5,3,4}\orcidlink{0000-0003-3701-5882}
          }

   \institute{Department of Physics and Astronomy,
              FI-20014 University of Turku, Finland\\
              \email{varpu.a.ahlberg@utu.fi} 
              \and
              Nordita, KTH Royal Institute of Technology and Stockholm University, Hannes Alfv\'ens v\"ag 12, SE-10691 Stockholm, Sweden
              \and
              Max Planck Institute for Astrophysics, Karl-Schwarzschild-Str. 1, D-85741 Garching, Germany
              \and
              Space Research Institute (IKI), Profsoyuznaya 84/32, Moscow 117997, Russia
              \and
              Rudolf Peierls Centre for Theoretical Physics, Department of Physics, University of Oxford, Clarendon Laboratory, Parks Rd, Oxford, OX1 3PU, United Kingdom
             }

   \date{Received XXXX / Accepted XXXX}
 
  \abstract
   {
   Super-Eddington accretion onto compact objects is expected to produce optically thick outflows with a funnel-shaped cavity that may collimate the emission.  
   At inclinations higher than the grazing angle of the funnel, the central source is obscured.
   Accordingly, the observed emission is dominated by scattered and reflected radiation, which can therefore be strongly polarized.
   The detection of strong X-ray polarization in the Galactic X-ray binary Cygnus X-3 provides the first direct probe of this geometry.
   In this work, we present a systematic study of the inclination-dependent radiative signatures of such systems using a combination of semi-analytical methods and Monte Carlo simulations.  
   Our treatment explicitly accounts for multiple scatterings and demonstrates that both the polarization degree and the degree of collimation are highly sensitive to the albedo of the funnel surface.
   We find that a low albedo (significant absorption) is essential for producing high polarization, yet it simultaneously suppresses the collimation of the emission. 
   Conversely, a high-albedo medium (nearly pure scattering) can modestly collimate radiation, but at the cost of substantially reducing the polarization degree.
   We discuss our results in the context of Imaging X-ray Polarimetry Explorer observations of Cygnus X-3 and propose a physical scenario for its spectral state transitions, considering a combination of reflection from the funnel surface and scattering by a diffuse medium above the funnel.
   Our model provides a general framework for interpreting X-ray polarimetric signatures of obscured accretors.
   } 


   \maketitle

\section{Introduction}

Super-Eddington accretion is a process of fundamental significance across cosmological timescales, thought to play a key role in phenomena ranging from the rapid growth of supermassive black holes in the early Universe to the extraordinarily bright emission of ultraluminous X-ray sources (ULXs).
ULXs are X-ray binaries (XRBs) whose inferred isotropic luminosities appear to surpass the Eddington limit for ten solar-mass black holes \citep[for reviews, see e.g.,][]{Kaaret2017,King2023}.
Their extreme apparent luminosities may arise from geometric beaming, i.e., the collimation of the central source radiation by a funnel-shaped cavity within the otherwise optically thick accretion flow or outflow \citep{King2001}.
Such funnel-type accretion geometry is expected to form in supercritical accretion disks, which either become advection-dominated \citep[``slim disks'',][]{Abramowicz1988} or eject their excess mass supply through a radiatively driven, Compton-thick wind \citep{SS73,Poutanen2007}. 
In this scenario, the super-Eddington system observed at a high viewing angle, away from the funnel axis, is expected to be dim and obscured.

While there are no persistent face-on ULXs in the Milky Way,\footnote{The apparent luminosity of the transient ULX pulsar \mbox{Swift J0243.6+6124} exceeded the Eddington limit by a factor of 40 during its 2017 outburst \citep{Tsygankov2018}.} there is evidence that some Galactic XRBs may represent ULXs viewed at high inclinations.
Signs of strong obscuration have been identified in several systems, including \mbox{SS 433} \citep{Fabrika04}, \mbox{GRS 1915+105} \citep{Miller2020}, \mbox{V404 Cygni} \citep{Motta2017}, \mbox{V4641 Sagittarii} \citep{2002A&A...391.1013R}, and \mbox{Cygnus X-3} \citep[Cyg X-3 hereafter,][]{Hjalmarsdotter2008}.
During certain epochs, their X-ray spectra show remarkable similarities, exhibiting spectral features arising from Compton-thick material surrounding the central source \citep{Koljonen2020}.
Owing to their relative proximity, these systems are valuable laboratories for probing the geometry of the innermost accretion flow and thereby testing models of supercritical accretion. 

Polarimetry provides a powerful diagnostic of the geometry of accreting systems. 
With the recent launch of the Imaging X-ray Polarimetry Explorer \citep[IXPE;][]{Weisskopf2022}, it has become possible to measure the polarization of X-ray emission from compact objects, opening a new avenue for probing the structure of their innermost regions.
The Galactic XRB \mbox{Cyg X-3}, long suspected to be strongly obscured, became the first source of its kind to be observed with IXPE.
The system is known to continuously switch between spectral states, characterized by distinct X-ray and radio properties \citep{Szostek2008, Hjalmarsdotter2009}.
Its typical hard X-ray, quiescent radio state has a spectrum consistent with Compton reflection from a cool medium or strong intrinsic absorption \citep{Hjalmarsdotter2008}.
The hard state is followed by an intermediate state, characterized by enhanced X-ray emission with similar yet softer spectra and minor radio flaring.
The intermediate state occasionally evolves further into a radio-quenched ultrasoft state with a blackbody spectrum, which can then develop into a major radio flaring state (the soft nonthermal X-ray state).
The nature of the compact object in \mbox{Cyg X-3}, either a black hole or a neutron star, is unclear due to large uncertainties in the mass measurements, although it has been argued that it is a low-mass black hole \citep{Zdziarski2013, Koljonen2017}.

The first IXPE observations of \mbox{Cyg X-3} revealed that its X-ray emission in the 2--8 keV range is highly polarized \citep{Veledina2024}, with an orbital phase-averaged polarization degree (PD) of $20.6\pm0.3\%$ in the hard X-ray spectral state \citep[the quiescent radio state;][]{Szostek2008}.
The polarization direction, defined by the polarization angle (PA), is orthogonal to the projected axis of the previously detected radio ejections \citep{Marti2000, Miller2004, Yang2023}.
The high PD, together with its energy dependence and polarization direction, suggests that the X-ray continuum is produced by reflected emission, while the primary emission is hidden behind an optically thick envelope containing a funnel-shaped cavity.
Modeling of the polarization signatures provided tight constraints on the funnel opening angle, which was found to be $\lesssim$15\degr.
Such a geometry is characteristic of super-Eddington accretion, making \mbox{Cyg X-3} the first source in which the funnel-type accretion geometry has been probed directly.

The hard-state accretion geometry of \mbox{Cyg X-3} appears remarkably stable, as subsequent observations taken a year later found very similar polarization signatures \citep{Mikusincova2025}.
However, the polarization shows clear variations both with the orbital period and with spectral transitions \citep{Veledina2024,Ahlberg2025,Mikusincova2025}.
IXPE observations of \mbox{Cyg X-3} covered its intermediate and ultrasoft states, which both show a PD of roughly $10\%$ with the same orientation as the hard state \citep{Veledina2024,Veledina2024soft}.
Substantial changes in polarization properties suggest a change in the accretion structure, yet the persistently high and energy-independent PD indicates that the funnel-type geometry is continually present. 

Scattering funnels and obscuring tori have been previously established as key elements of the active galactic nucleus unification scheme \citep{AntonucciMiller1985,Pogge1989,Netzer2015}.
Imprints of a funnel-type geometry on spectral and polarimetric signatures have been studied both in the context of extragalactic sources and small-scale, Galactic systems, where the role of the obscuring torus could be replaced by an optically thick outflow \citep{Ghisellini1994,Ratheesh2021}.
The same geometry can be extended to cases with small opening angles, where the funnel can collimate radiation, explaining the remarkable brightness of ULXs \citep{King2001,Dauser2017}.
Recently, motivated by new X-ray polarimetric observations of \mbox{Cyg X-3}, both analytical and Monte Carlo (MC) approaches have been used to explore the polarization properties of funnel-like outflows \citep{Veledina2024,Podgorny2023,Podgorny2025}.
These studies explored the energy dependence of polarization and highlighted the key dependence of resulting polarization signatures on the ionization state of the outflow and the angular dependence of the incident emission.
They found a relative robustness of polarization signatures against the details of the funnel shape.
However, previous works have largely focused on hard-state configurations and, importantly, have relied on the single-scattering or single-reflection approximations.

In this work, we developed a more general model of radiative transfer in super-Eddington outflows by relaxing this assumption and explicitly accounting for multiple scatterings.
We employed both semi-analytical methods and direct MC simulations to examine how the funnel geometry influences the polarimetric properties, particularly focusing on multiple scattering and reflection processes under the assumption of monochromatic scattering.
We carried out a systematic parameter study and compared our results with observations of \mbox{Cyg X-3}.
Our approach provides a comprehensive framework for interpreting polarimetric signatures of obscured super-Eddington sources and highlights the qualitative and quantitative differences that arise once multiple scatterings are properly included.

In Sect. \ref{sect:models}, we describe the analytical approximation and the MC code for electron scattering in a conical funnel.
In Sect. \ref{sect:results}, we present the results of our parameter study.
In Sect.~\ref{sect:discussion}, we discuss the implications for obscured super-Eddington XRBs and introduce a new physical scenario for the spectral transitions in \mbox{Cyg X-3}.
Finally, in Sect. \ref{sect:summary}, we summarize our main findings.

\section{Models} \label{sect:models}

\subsection{Geometry} \label{sect:geom}

\begin{figure}
\begin{center}
\includegraphics[width=0.7\linewidth]{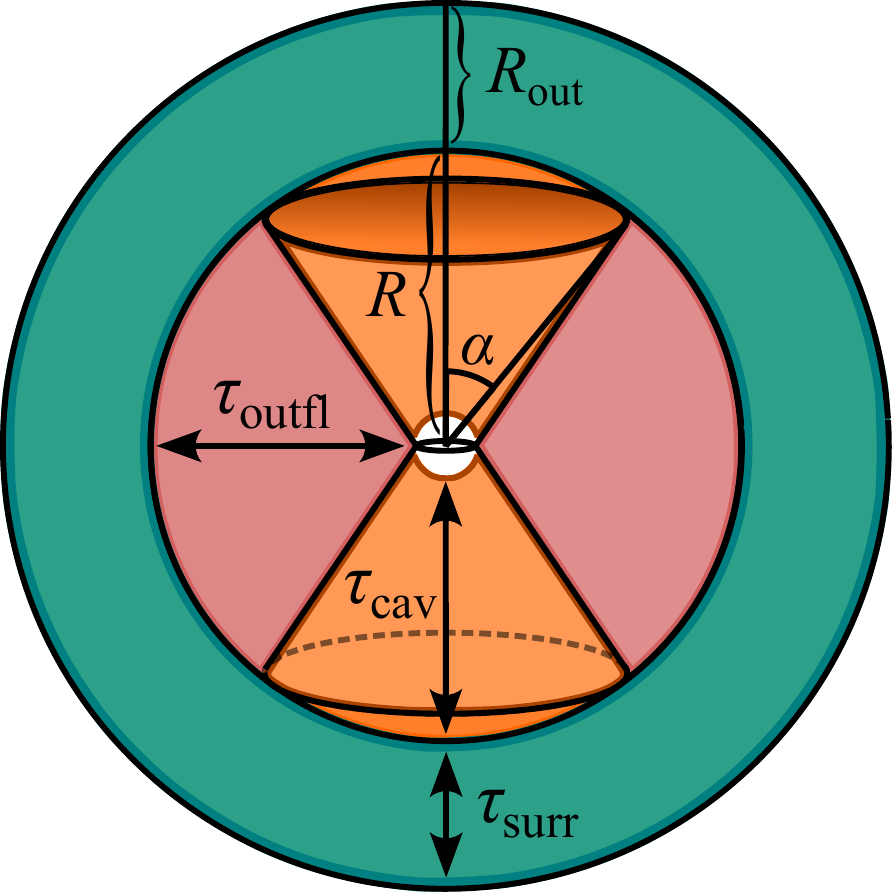}
\end{center}
\caption{Illustration of the funnel geometry and the three scattering regions: the outflow (brown), the inside of the funnel cavity (orange), and the surrounding area beyond the outflow (teal). The innermost spherical region (white) is devoid of gas.}
\label{fig:conegeom}
\end{figure}

In our models, we considered a compact object surrounded by an obscuring outflow.
We characterized the shape of the funnel cavity as a double conical frustum with two parameters: the distance from the central source to the edge of the funnel, $R$, and the grazing angle of the funnel, $\alpha$.
We expressed $R$ and all other distances in dimensionless units as multiples of the frustum base radius, which we expect to be on the order of several Schwarzschild radii \citep{Poutanen2007}.
With these, the height of the funnel is $h=R \cos \alpha$, its maximal cylindrical radius is $\rho_\mathrm{max}=R\sin\alpha$, and its half-opening angle is $\tan\zeta = (\rho_\mathrm{max}-1)/h$.
The outflow forms a toroid encircling the frustum cavity with a spherical outer boundary of radius $R$.
This geometry is illustrated in Fig. \ref{fig:conegeom}.
We divided the outflow and its surroundings into three regions: the outflow itself, the inside of the funnel cavity up to a radius of $R$, and the surrounding area at radii beyond $R$.
The surrounding region is limited to a maximal radius of $R+R_\mathrm{out}$, which we set at $R_\mathrm{out} = 0.5R$.
Since the medium directly above the funnel cavity is exposed to the direct emission, only this part of the surrounding region significantly contributes to the scattered flux.
As a simplification, we kept the innermost spherical region within the frustum base radius empty of gas to exclude an accretion disk and interactions very close to the central source.
As a consequence, the light is allowed to pass through the hole at the center of the cavity.

We treated the source of the incident emission as an isotropic and unpolarized point source of luminosity $L$ located at the center of the funnel.
The observer is situated at an inclination angle of $i$ from the cone axis (vertical direction in Fig. \ref{fig:conegeom}) and a distance of $D$ from the source.
The direct, unobscured flux from the source would therefore be $F_0=L/\left(4\pi D^2\right)$.

We described linear polarization with the Stokes fluxes, $F_I$ and $F_Q$, from which the PD can be calculated as $|F_Q|/F_I$.
For the polarization basis, we used the projection of the cone axis on the plane of the sky (see Appendix \ref{sect:reflection}).
The Stokes $U$ of the total emission is zero in our models given the assumed axial symmetry of the scattering medium.
To simultaneously display both the PD and PA, we used the normalized Stokes parameter $q=F_Q/F_I$.
In this notation, the polarization is aligned with the cone axis when $q$ is positive (PA=$0^\circ$), and perpendicular to the axis when it is negative (PA=$90^\circ$). 

We gave the outflow a constant single-scattering albedo, $\lambda=\sigma_\mathrm{sc}/(\sigma_\mathrm{abs} + \sigma_\mathrm{sc})$, which is the ratio of the scattering cross section to the sum of the absorption and scattering cross sections.
We expect the optically thin gas to be fully ionized if the source luminosity surpasses the Eddington limit for solar mass objects, $L\gtrsim10^{38}~\mathrm{erg/s}$, as the ionization parameter is $\xi = L/(r^2 n)\gtrsim10^3$ out to a large distance of $r\lesssim10^{11}~\mathrm{cm}$ for electron number densities of $n\lesssim10^{13}~\mathrm{cm}^{-3}$ \citep[realistic value for the WR wind in Cyg X-3 near the compact object;][]{Antokhin22}.
Therefore, we only considered an albedo of unity for the surrounding matter and the matter inside the cavity.
We assumed that the three regions have uniform electron number densities, $n_\mathrm{outfl}$, $n_\mathrm{cav}$, and $n_\mathrm{surr}$, which we parameterized by their optical depths: the outflow has a horizontal Thomson scattering optical depth of $\tau_\mathrm{outfl} = \sigma_\mathrm{sc}n_\mathrm{outfl} (R-1)$, and the inner and outer regions have vertical Thomson scattering optical depths of $\tau_\mathrm{cav}=\sigma_\mathrm{sc} n_\mathrm{cav} (R-1)$ and $\tau_\mathrm{surr}=\sigma_\mathrm{sc}n_\mathrm{surr}  R_\mathrm{out}$, respectively (see Fig. \ref{fig:conegeom}).

\subsection{Analytical models}

An analytical single-scattering model explaining the hard state polarization of \mbox{Cyg X-3} with reflection from the optically thick funnel surface was presented in \citet{Veledina2024}.
In this scenario, there is no gas above or inside the cavity.
In contrast, the lower PD in the intermediate and ultrasoft states was analytically modeled in \citet{Veledina2024soft} with single scattering from an optically thin medium above the funnel cavity.
In this case, the funnel-reflected flux is assumed to be subdominant to the emission scattered by the optically thin gas.
To estimate how the PD changes as a function of the optical depth of the medium above the funnel, we combined these two analytical models.
For consistency, we refer to the light scattered by the optically thick funnel surface as the ``reflected'' emission (subscript refl) and the light scattered by the optically thin medium as the ``scattered'' emission (subscript sc).

The Stokes flux of the radiation reflected by the funnel surface under the single-scattering approximation is
\begin{equation} \label{eq:funnelreflection}
    \left( \begin{matrix}
        F_{I, \mathrm{refl}} \\
        F_{Q, \mathrm{refl}}
    \end{matrix} \right) = 
    F_0\frac{3\lambda}{16\pi}
    \int\limits_{-h}^{+h}  \int\limits_{0}^{2\pi} \frac{\rho}{z^2 + \rho^2}
 \left( \begin{matrix} 
      1+\mu^2\\
      (1-\mu^2) \cos 2 \chi
    \end{matrix} \right)
    \frac{\eta \eta_0}{\eta + \eta_0}
    \, \frac{\mathrm{d}\phi\,\mathrm{d}z}{\cos \zeta},
\end{equation}
where $z$ is the height along the funnel axis, $\rho$ and $\phi$ are the cylindrical radius and azimuth of the funnel surface, respectively, $\mu$ is the cosine of the scattering angle, $\eta_0$ and $\eta$ are cosines of the incidence and outgoing angles, and $\chi$ is the PA.
The detailed description of the surface integral is given in Appendix \ref{sect:reflection}.

\begin{figure}
\begin{center}
\includegraphics[width=\linewidth]{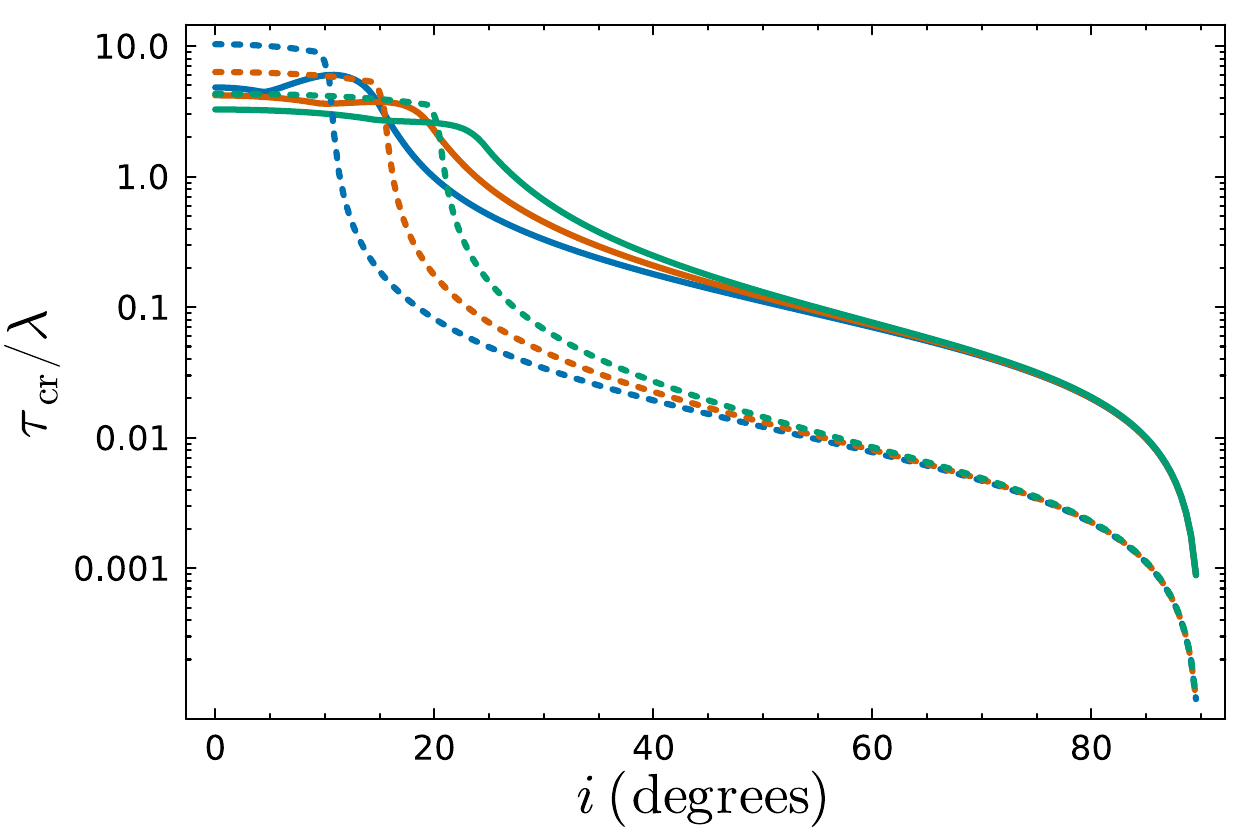}
\end{center}
\caption{Ratio of the critical optical $\tau_\mathrm{surr} = \tau_\mathrm{cr}$ depth of gas above the funnel to the single-scattering albedo of the outflow as a function of inclination for funnel grazing angles of $\alpha=10\degr$ (blue), $\alpha=15\degr$ (orange), and $\alpha=20\degr$ (green), and funnel lengths of $R=10$ (solid) and $R=100$ (dashed). For this value of $\tau_\mathrm{surr}$, the flux reflected by the funnel surface and the flux scattered from above the funnel cavity are equal.}
\label{fig:criticaldepth}
\end{figure}

We modeled the geometry of the medium above the funnel as the spherical shell described in Sect. \ref{sect:geom}, although we only considered the section immediately above the funnel opening, since the rest is not directly illuminated.
This treatment is similar to the model employed in \citet{Veledina2024soft} for the scattering medium within the funnel.
In spherical coordinates, the electron number density of the region is
\begin{equation}
    n_\mathrm{surr}(r,\theta) = \begin{cases}
    n_\mathrm{surr}, & \mathrm{if} \ \  R\leq r\leq R+R_\mathrm{out} ~~\mathrm{and} \ \ \theta\leq\alpha,   \\
    0,  & \mathrm{otherwise,}
    \end{cases}
\end{equation}
where $r$ is the radial distance from the central source and $\theta$ is the colatitude measured from the cone axis.
We assumed that the region is sufficiently optically thin to transmit the single-scattered emission without attenuation.
A phenomenological description for the attenuation effect was considered in \citet{Chaurasia2025}.
For unpolarized incident light from an isotropic point source, the scattered flux can be found analytically with Eqs. (5--7) from \citet{Brown1978}:
\begin{equation} \label{eq:scatterflux}
    \left( \begin{matrix}
        F_{I, \mathrm{sc}} \\
        F_{Q,\mathrm{sc}}
    \end{matrix} \right) = F_0\frac{3\tau_\mathrm{surr}}{16}  \left(\begin{matrix}
    \frac{8}{3} - 2 \cos \alpha \left[1 + \frac{1}{3} \cos^2 \alpha \right] - \sin^2 i \cos \alpha \sin^2 \alpha \\
    - \sin^2 i \cos \alpha \sin^2 \alpha
        
    \end{matrix} \right).
\end{equation}
This model relies on the assumption that the medium above the funnel is fully visible at all inclinations, and thus neglects any possible obscuration by the outflow or the accretion disk.
We also assumed that the inside of the cavity is largely concealed and thus did not include scattering from a medium filling the cavity in the analytical model.

To find the optical depth where the flux scattered from above the funnel begins to dominate the funnel reflection, we calculated the critical value $\tau_\mathrm{surr}=\tau_\mathrm{cr}$ for which $F_{I,\mathrm{refl}}/F_{I,\mathrm{sc}} = 1$.
From Eqs. (\ref{eq:funnelreflection}) and (\ref{eq:scatterflux}), we thus find $\tau_\mathrm{cr}/\lambda$ as a function of $i$.
Figure \ref{fig:criticaldepth} shows how $\tau_\mathrm{cr}/\lambda$ depends on the observer inclination for different values of $R$ and $\alpha$.
The critical optical depth decreases rapidly with increasing inclination regardless of the funnel grazing angle.
At low inclinations, however, the geometrical parameters have a significant impact on the critical optical depth.
For example, for an albedo of $\lambda = 0.15$, the critical optical depth at $i=30^\circ$ is about 0.05 with $R=10$ and $\alpha=10^\circ$, and with $R=10$ and $\alpha=20^\circ$, it is doubled to $\tau_\mathrm{cr} \sim 0.1$.
At larger values of $R$, the critical depth decreases more rapidly with inclination, as less light is reflected toward the sides.

\subsection{Monte Carlo simulations}

The analytical model describing the reflection from a semi-infinite funnel does not account for multiple scatterings within a finite-opacity outflow, secondary scatterings from a medium inside or above the funnel, nor for multiple reflections across the funnel cavity.
To test the impact of these effects on the polarization, we employed MC simulations of the radiative transfer.

The MC code we used is a variation of the family of radiative transfer codes developed for modeling X-ray illumination of neutral or ionized media  described in \citet{Churazov2017} and references therein.
In particular, it has been employed to simulate the reflection off optically thick disk funnel walls and optically thin disk wind in SS~433 \citep{Medvedev2018}.
To make the comparison between the analytical and MC results straightforward, we introduced the following simplifications: (i) we used the dipole scattering phase function, (ii) we kept the simulations monochromatic, i.e., the energy does not change after scattering, and (iii) we parameterized the opacities only with $\lambda$ and the scattering optical depths rather than calculating them from the chemical composition and the ionization state of the medium.
These approximations correspond to the analytic single-scattering model of a semi-infinite funnel employed by \citet{Veledina2024} when only single-scattered photons are included and the outflow optical depth is very large.

Since we modeled the polarization rather than the full spectrum, the only relevant energy-dependent feature of the radiative transfer problem is the single-scattering albedo.
As long as the recoil effect is insignificant, the dependence of the reflection on $\lambda$ can therefore mimic the energy dependence.
The monochromatic treatment does not account for the thermal emission of the funnel surface, which would add an unpolarized component to the flux at low energies.
We also did not account for the outflowing velocity of the medium to simplify our analysis, since our preliminary results were qualitatively similar regardless of mildly relativistic velocities.

The densities of the absorbing and scattering media (see Fig. \ref{fig:conegeom}) were stored in an evenly spaced 2D cylindrical grid, which represented the axially symmetric 3D geometry in $\sqrt{x^2 + y^2}$-$z$ coordinates.
We note that the photon trajectories are analytically solvable in our simplified model, but we opted for a gridded approach for future flexibility.
The optical depth in the photon propagation direction was integrated numerically along the grid and used to determine the escape fraction and the location of the next interaction.
The statistical weight of each photon packet was then updated to account for the escaped and absorbed fractions.
The weights of the photon packets exiting toward the observer were tallied into 100 equally spaced bins of $|\cos i|$.
We traced at least $2.4\times10^5$ photons during each simulation, using 10 times more photons in Sect. \ref{sect:surfacereflection} to more accurately compare with the analytical model.

\begin{figure*}
\sidecaption
\includegraphics[width=12cm]{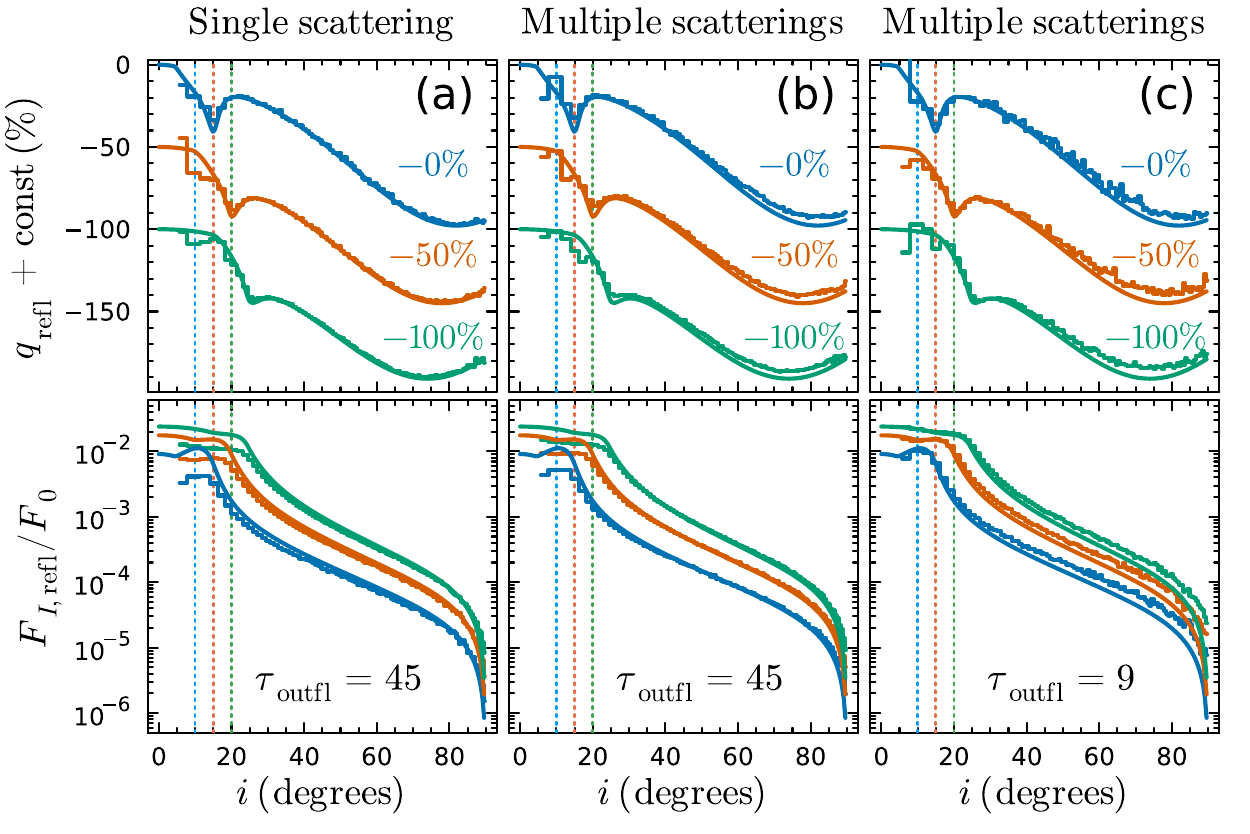}
\caption{Comparison of the normalized Stokes $q$ (top, PD is $|q|$) and the flux (bottom) of light reflected from the funnel surface between the analytical (line) and MC (histogram) methods.
The flux is scaled by the unobscured incident flux, $F_0$.
Panels (a) have a large horizontal scattering optical depth of $\tau_\mathrm{outfl}=45$ while including only single-scattered photons, whereas panels (b) show the same case while including multiple scatterings in the MC simulations.
Panels (c) also include multiple scatterings, but for a less opaque funnel with $\tau_\mathrm{outfl}=9$.
The funnel has a length of $R=10$ and a small single-scattering albedo of $\lambda=1/6$, with different funnel grazing angles of $\alpha=10\degr$ (blue), $\alpha=15\degr$ (orange), and $\alpha=20\degr$ (green).
We added multiples of $-50\%$ to the different $q$ results to space them vertically. 
The vertical dotted lines show the inclination $i=\alpha$ at which the central source becomes visible.}
\label{fig:analytical_comparison}
\end{figure*} 

\section{Results} \label{sect:results}

\subsection{Reflection from the funnel surface} \label{sect:surfacereflection}

We first examined the reflection from the funnel surface under the optically thick single-scattering assumption used by the analytical model to explore the secondary effects present in the MC simulations.
We set the funnel length to $R=10$ and used three different funnel grazing angles of $\alpha=10^\circ$, $15^\circ$, and $20^\circ$.
To reproduce a fully opaque funnel surface featuring mostly single scattering with the MC code, we set the horizontal scattering optical depth of the outflow to $\tau_\mathrm{outfl}=45$ (5 for each $\Delta R=1$) and the ratio of the absorption and scattering cross sections to $\sigma_\mathrm{abs}/\sigma_\mathrm{sc}=5$, resulting in a single-scattering albedo of $\lambda=1/6$.
This fiducial value for the single-scattering albedo is approximately that of neutral gas with solar abundances within the IXPE energy range \citep[see, e.g.,][]{Basko1974}.
We additionally tested a less optically thick outflow with $\tau_\mathrm{outfl}=9$ (1 for each $\Delta R=1$) to study the impact of a finite opacity.

Figure \ref{fig:analytical_comparison} shows the results of the MC simulations.
As expected, the PD of the single-scattered emission for $\tau_\mathrm{outfl}=45$ is identical to the analytical model.
However, the reflected fluxes of the MC and analytical models diverge at low inclinations, with a large portion of the flux missing at $i<20\degr$ in the higher optical depth simulation.
This is caused by our parameterization of the scattering medium in a 2D grid.
Chiefly, if the integration step is too large, some photon packets may penetrate erroneously deep into the funnel and become absorbed before escaping.
Errors caused by the grid can be resolved by increasing its resolution, which is computationally expensive.
The $\tau_\mathrm{outfl}=9$ simulation shows no reduction of the flux compared to the analytical model, as the optical depth spanned by each integration step is smaller.

In both the analytical and MC models, the PD has two local maxima that occur when light reflected at an angle of $\sim$$90^\circ$ is seen.
The primary maximum in PD occurs at high inclinations when only the upper rim of the funnel surface is visible, and the secondary maximum appears at low inclinations when light reflected from close to the funnel base comes into view.
At high inclinations, however, the PD is slightly reduced in the simulations that accounted for multiple scatterings.
This occurs because more photons escape along the surface normal of the funnel.
Therefore, multiple scatterings contribute more when the line of sight is perpendicular to the funnel surface at $i\sim90^\circ-\alpha$. 

The flux at high inclinations is increased in the $\tau_\mathrm{outfl}=9$ case, since a small section of the funnel surface is transparent.
However, this does not significantly alter the PD.
To illustrate why this happens, we calculated the horizontal scattering optical depth through the outflow as a function of the elevation, $z$:
\begin{equation}
    \tau_z =\frac{\tau_\mathrm{outfl} }{R-1} \left[ R\sqrt{1-\frac{z^2}{h^2}\cos^2 \alpha} - (R\sin\alpha -1)\frac{z}{h} - 1 \right],
\end{equation}
where $h=R \cos \alpha$ is the height of the funnel.
Here, the scattering optical depth through the outflow is less than unity in a narrow region above $z/h>0.95$ for $\alpha \leq 20^\circ$ and $R=10$. 
Even though the optically thin portion of the outflow is too small to meaningfully scatter the emission, it effectively increases the visible area of the reflecting surface, and thus increases the flux.

\begin{figure*}
\sidecaption
\includegraphics[width=12cm]{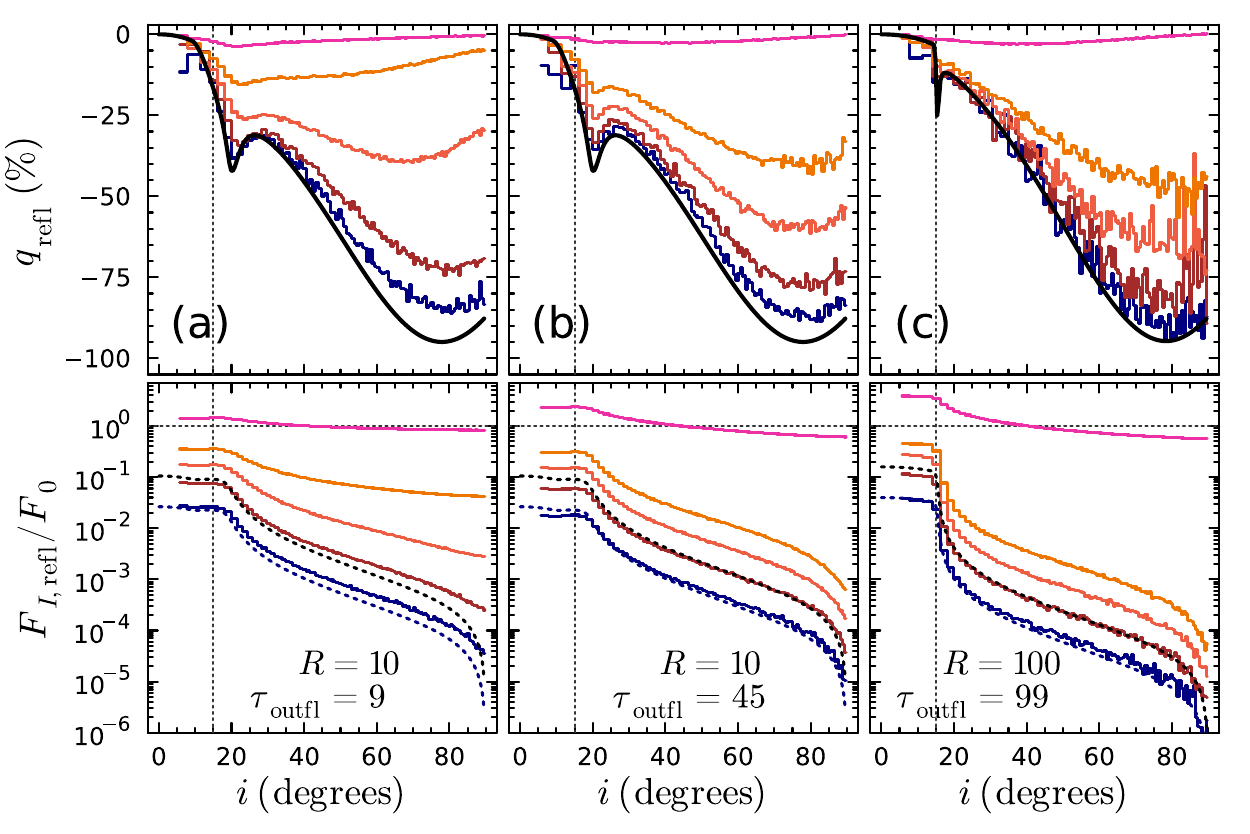}
\caption{Normalized Stokes $q$ (PD is $|q|$; top), and scaled reflected flux (bottom) from the funnel surface with single-scattering albedos $\lambda=0.25$, 0.5, 0.75, 0.9, and 1.0 (blue, brown, orange, yellow, and magenta, respectively) with $\alpha=15\degr$ for different funnel lengths and horizontal scattering optical depths: $R= 10$ with $\tau_\mathrm{outfl} = 9$ (a), $R=10$ with $\tau_\mathrm{outfl}=45$ (b), and $R=100$ with $\tau_\mathrm{outfl}=99$ (c).
For comparison, the black $q$ line and dashed flux lines show the analytical single-scattering model with albedos of 0.25 (blue) and 1.0 (black). To highlight the weakness of the geometric beaming, the dotted gray lines show the flux from the unobscured central source, $F_0$, and the inclination at which the central source becomes visible, $i=\alpha$.}
\label{fig:albedo}
\end{figure*}

\subsection{Reflection with increased albedo} \label{sect:albedo}

Next, we considered the effects of an increased albedo for the reflection from the funnel surface.
This increases the number of scatterings a photon can survive prior to being absorbed, which is expected if the outflow is highly ionized.
We performed simulations with different values for $\lambda$, including a purely scattering case ($\lambda = 1.0$), at a fixed funnel grazing angle of $\alpha=15\degr$.
We investigated the case of $R=10$ for both $\tau_\mathrm{outfl}=9$ and $\tau_\mathrm{outfl}=45$, but also included a more extended funnel with $R=100$ and $\tau_\mathrm{outfl}=99$.

Figure \ref{fig:albedo} shows the results for $\lambda=0.25$, $0.5$, $0.75$, $0.9$, and $1.0$.
Intriguingly, the reflection leads to appreciable beaming only when the albedo approaches unity.
This is a consequence of the typically large number of scatterings that photons emitted toward the funnel surface experience before escaping the funnel.
To estimate the number of reflections across the funnel cavity, we made an MC toy model of the geometry with a perfectly reflecting Lambertian surface.
We found that for $R=10$ and $\alpha=15^\circ$, the average number of reflections is $\approx$12.
When combined with multiple interactions within the outflow, the photons may undergo dozens of interactions prior to escape.
Assuming that the photons have a probability of $\lambda^N$ of surviving $N$ interactions without being absorbed, the expected number of interactions before absorption is $\left(\sum\limits_{N=1}^{\infty} N\lambda^N \right)/  \left(\sum\limits_{N=1}^{\infty} \lambda^N \right) = (1-\lambda)^{-1}$.
Therefore, the majority of the emission is absorbed unless the albedo is $\lambda >0.9$.

At inclinations of $\lesssim45\degr$ and albedos of $\lesssim0.5$, the PD in the MC simulations is smaller by only a few percent compared to the analytical model.
The analytical and MC models also produce comparable reflected fluxes at $\lambda=0.25$ (Fig. \ref{fig:albedo}, solid and dashed blue lines).
However, the reflected flux in the MC model depends nonlinearly on $\lambda$ when multiple reflections across the cavity become significant.
The flux increases with the albedo more rapidly in the MC simulations than in the analytical model, particularly between $\lambda=0.9$ and $\lambda=1.0$ (orange and magenta lines).

The PD in the MC simulations is more notably reduced compared to the analytical model at high inclinations, as some of the emission propagates through the outflow without being absorbed.
This ‘‘leakage’’ through the funnel surface is also apparent by the greater relative increase in the flux at high inclinations (see Fig. \ref{fig:albedo}a at $i>45^\circ$).
This occurs even though the outflow is opaque to scattering, since a fraction of the photons become trapped at large optical depths and can therefore escape through the outer boundary of the outflow.

When $\lambda=1.0$ (magenta lines of Fig. \ref{fig:albedo}), the massively increased number of scatterings reduces the PD to $\lesssim5\%$ across all our simulations.
The purely reflecting funnel surface can collimate the emission, particularly when $\tau_\mathrm{outfl}$ is large, but the effect is diminished due to the flux leaked through the outer boundary of the outflow.
Strikingly, the flux is distributed nearly isotropically for the least optically thick case of $\tau_\mathrm{outfl}=9$.
In the ideal scenario for geometric beaming, the flux is enhanced by a factor of $ (1-\cos\alpha)^{-1}\approx29$ that is inversely proportional to the solid angle subtended by the funnel.
However, even in our highest optical depth simulation with $\tau_\mathrm{outfl}=99$, the reflected flux is only a factor of $\sim$4 greater than the direct flux when viewed at an inclination of $i=0^\circ$.
Even for a truly semi-infinite funnel surface, we expect that the flux boost would still be diminished, since some of the light is reflected toward the sides.

\subsection{Scattering above and inside the funnel cavity}

Next, we considered the funnel filled with an optically thin scattering medium.
The increased density of gas enhances the number of photons that are scattered directly toward the observer, increasing the flux.
This also reduces the PD, since on average, the medium scatters the light at smaller angles ($\sim$$i$) than the funnel surface ($\sim$$i+\alpha$).
To find the optical depth when scattering from the diffuse medium begins to dominate the polarization, we included a purely scattering gas for three different cases: (i) above the funnel, (ii) inside the funnel cavity, (iii) both above and inside the funnel.
Similar to the cases considered above, we fixed the funnel length to $R=10$ and tested different funnel grazing angles of $\alpha=10^\circ$, $15^\circ$, and $20^\circ$.
We set the outflow optical depth to $\tau_\mathrm{outfl}=18$ to reduce the amount of erroneous absorption caused by a large integration step.
To focus on single scattering, we set the albedo of the outflow to $\lambda=1/6$.

\begin{figure*}
\sidecaption
\includegraphics[width=12cm]{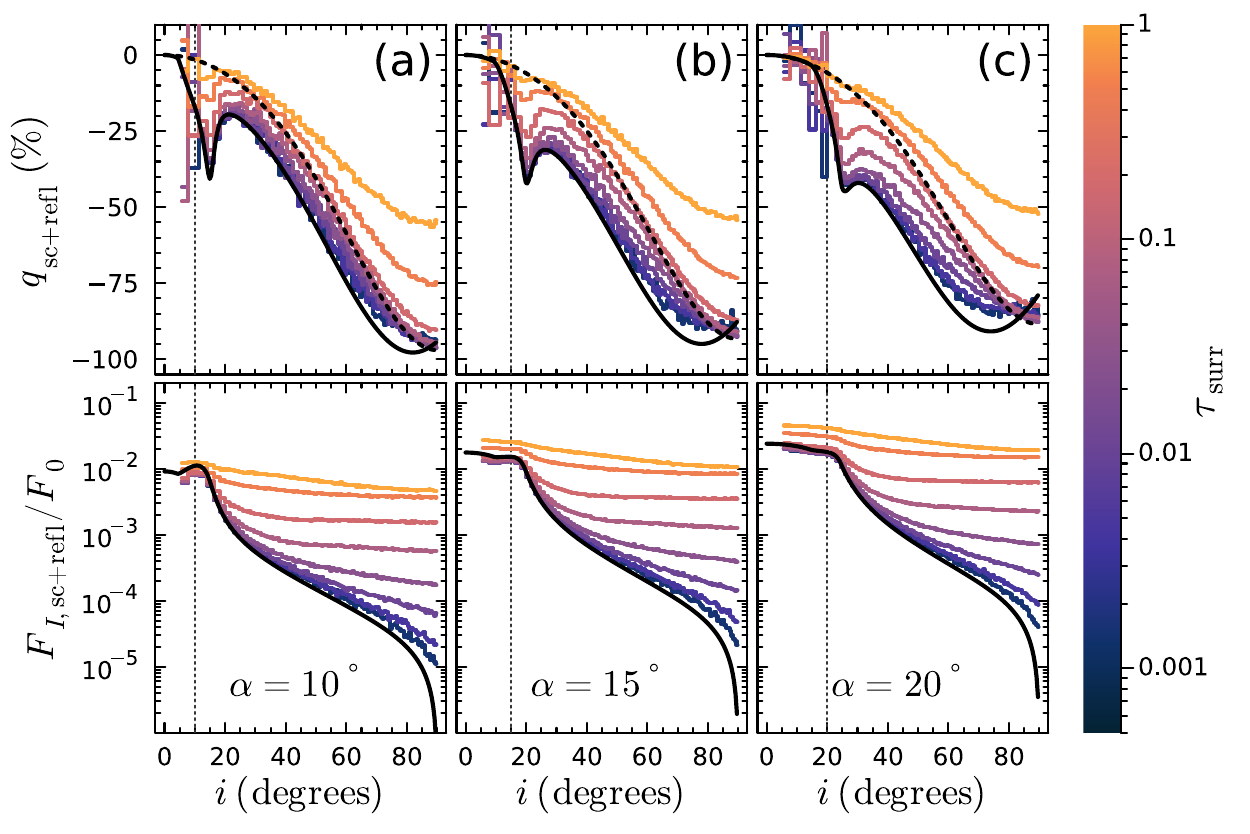}
\caption{Normalized Stokes $q$ (PD is $|q|$; top) and scaled flux (bottom) of the reflected and scattered emission at 14 different vertical optical depths $\tau_\mathrm{surr}$ of gas located above the funnel (outer region in Fig. \ref{fig:conegeom}) with $R=10$ and three different funnel grazing angles: $\alpha=10\degr$ (a), $\alpha=15\degr$ (b), and $\alpha=20\degr$ (c).
The horizontal scattering optical depth of the funnel is $\tau_\mathrm{outfl}=18$ and its single-scattering albedo is $\lambda=1/6$, and $\lambda_\mathrm{cav}=0$.
Ranging from dark purple to orange, the optical depths are $\tau_{\rm surr}=0.0005$, 0.0015, 0.005, 0.015, 0.05, 0.15, 0.5, and 1.0.
The solid and dashed black lines represent the analytical models for reflection from the funnel surface and single scattering from the medium above the funnel, respectively. }
\label{fig:I_PD_tau}
\end{figure*} 

\begin{figure}
\begin{center}
\includegraphics[width=\linewidth]{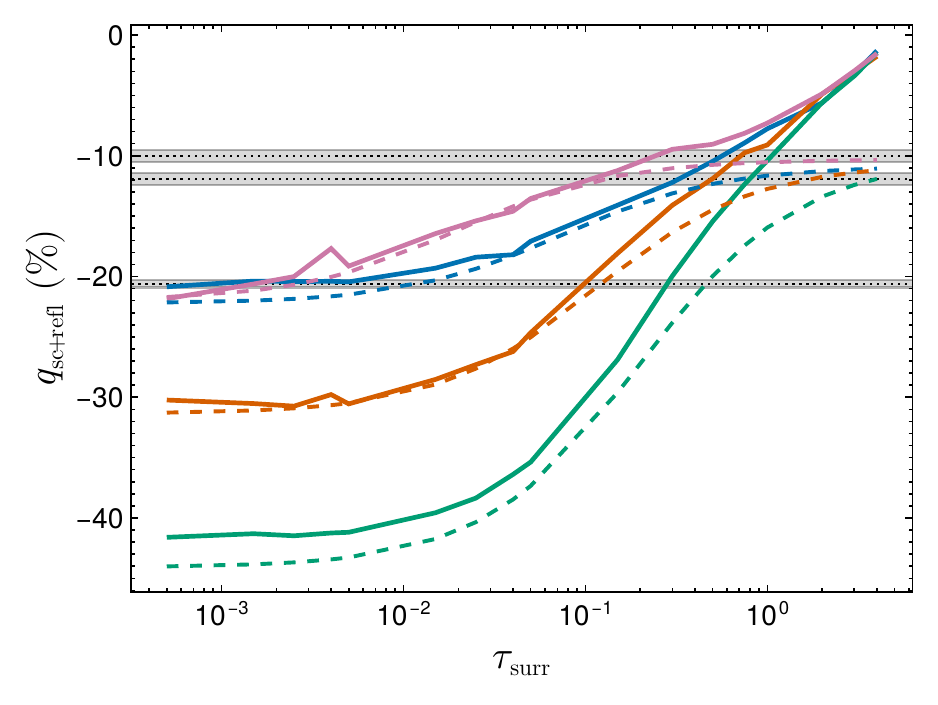}
\end{center}
\caption{Normalized Stokes $q$ (PD is $|q|$) of the reflected and scattered light as a function of $\tau_\mathrm{surr}$ averaged over an inclination range of 
[$25^\circ$--$28^\circ$]. 
The solid lines show the results of the MC simulation, while the dashed lines show the analytical approximation.
The funnel parameters are $\alpha=10\degr$ with $R=10$ (blue), $\alpha=15\degr$ with $R=10$ (orange), and $\alpha=20\degr$ with $R=10$ (green) and $R=100$ (lavender).
The other optical depths are $\tau_\mathrm{outfl}=18$ and $\tau_\mathrm{cav}=0$, and the single-scattering albedo is $\lambda=1/6$.
To compare with the polarization changes in \mbox{Cyg X-3}, the dotted black lines show the observed polarization in the hard ($20.6\pm0.3\%$) and intermediate/ultrasoft states ($10.0 \pm 0.5\%$, $11.9 \pm0.5\%$) with their 1$\sigma$ error bars (gray ribbons).
}
\label{fig:PD_tau}
\end{figure} 

\begin{figure*}
\sidecaption
\includegraphics[width=12cm]{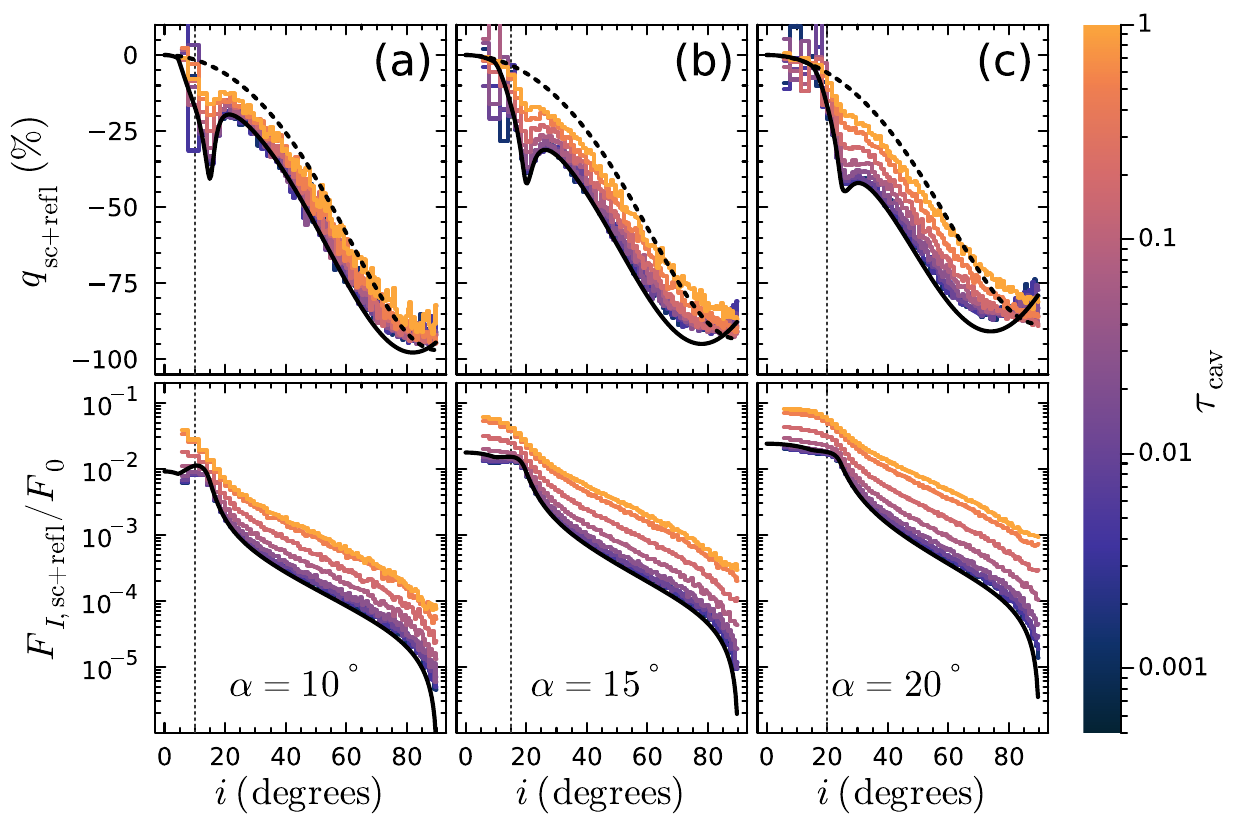}
\caption{Same as Fig. \ref{fig:I_PD_tau}, but with gas inside the funnel cavity (orange region in Fig. \ref{fig:conegeom}) and with $\tau_\mathrm{surr}=0$.}
\label{fig:I_PD_tau_in}
\end{figure*} 

Figure \ref{fig:I_PD_tau} shows the flux and polarization as a function of the vertical optical depth of gas above the funnel, $\tau_\mathrm{surr}$, with $\tau_\mathrm{cav}=0$.
The scattered flux becomes more evenly distributed across inclinations with increasing $\tau_\mathrm{surr}$, as the total angular distribution approaches the Thomson scattering phase function $\propto1+\mu^2$.
Since much of the total luminosity is emitted up toward the medium above the funnel, only a small optical depth of $\tau_\mathrm{surr}\gtrsim0.01$ is necessary to noticeably reduce the PD.

To illustrate how the polarization changes at a fixed inclination, Fig. \ref{fig:PD_tau} shows the PD as a function of $\tau_\mathrm{surr}$ averaged over three inclination bins spanning approximately $25^\circ$--$28^\circ$ to reduce the statistical noise.
We chose this inclination range to roughly correspond to the observed inclination of \mbox{Cyg X-3} \citep{Vilhu2009, Antokhin22}.
In addition to the previously considered cases, we tested a funnel with $R=100$ and $\alpha=20^\circ$.
We also calculated an analytical estimate of the PD as a function of $\tau_\mathrm{surr}$ by summing the Stokes fluxes of Eqs. (\ref{eq:scatterflux}) and (\ref{eq:funnelreflection}).
The PD of the MC simulations largely follows this analytical prediction at low optical depths, although multiple scatterings reduce the PD by a few percent.
At the critical optical depth estimated from Eqs. (\ref{eq:funnelreflection}) and (\ref{eq:scatterflux}) ($\tau_\mathrm{surr}\sim0.1$ for $R=10$ and $\tau_\mathrm{surr}\sim0.01$ for $R=100$), the PD is approximately halfway between the analytical values for the funnel surface reflection and the optically thin scattering.
The PD in the analytical model saturates roughly at a value of $10\%$ at large optical depths, whereas multiple scatterings significantly reduce the PD in the MC simulations when $\tau_\mathrm{surr}$ approaches unity.
This is particularly noticeable with wider funnel grazing angles due to our small value for $R_\mathrm{out}$, since some of the light travels a greater horizontal distance, and thus experiences a higher line-of-sight optical depth than $\tau_\mathrm{surr}$.
We predict that the MC model would also converge on a stable PD if the critical optical depth were lower.

Figure \ref{fig:I_PD_tau_in} shows the polarized reflection with optically thin gas inside the funnel cavity with different values of $\tau_\mathrm{cav}$, with $\tau_\mathrm{surr} = 0$.
The reduction of the PD with increasing optical depth is smaller in magnitude, yet qualitatively similar to the case with gas above the funnel.
This is influenced by our choice of spherical boundaries for the scattering regions, as the inner region protrudes slightly above the top of the funnel.
Unlike in the case of gas above the funnel, the flux remains strongly anisotropic and is greater than the flux reflected by an empty funnel across all inclinations.
The flux increase is a result of the thin medium scattering some of the light upward before it can reach the strongly absorbing funnel surface.

\begin{figure*}[ht!]
\sidecaption
\includegraphics[width=12cm]{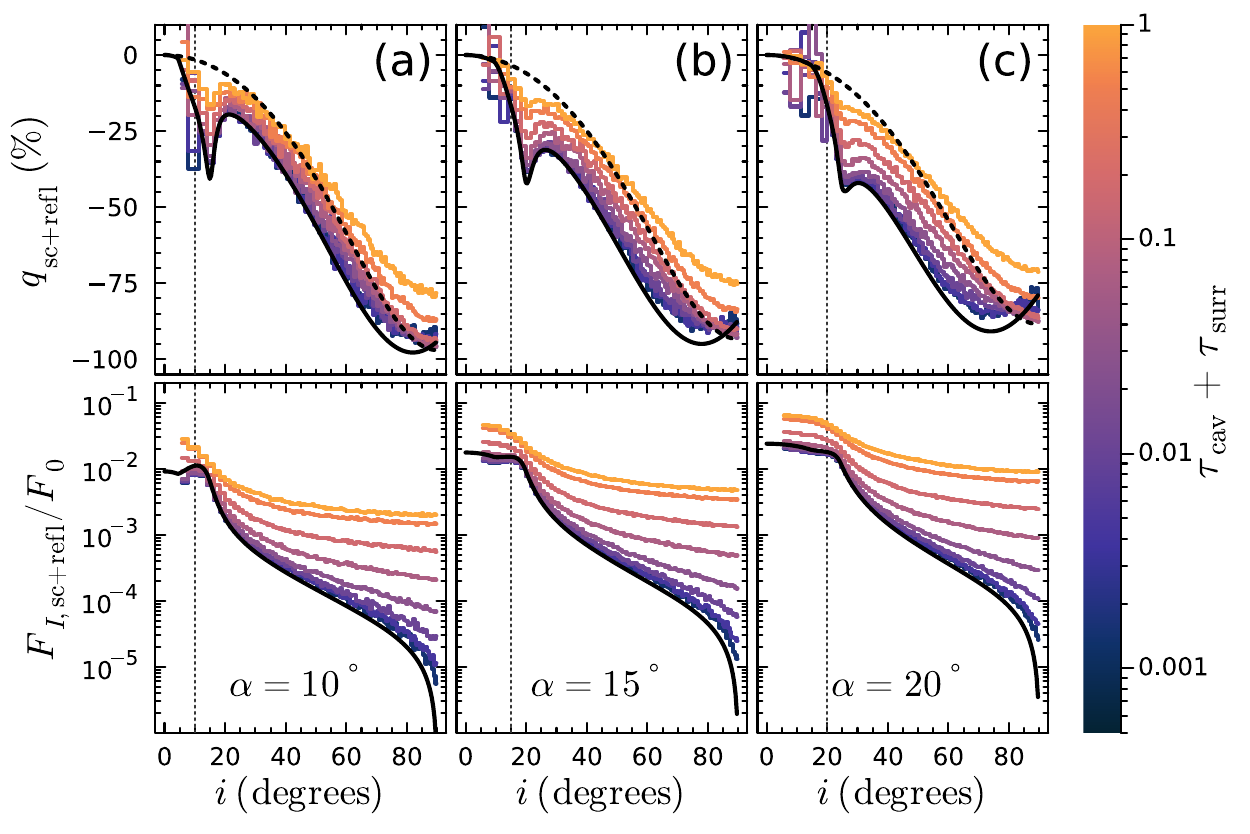}
\caption{Same as Fig. \ref{fig:I_PD_tau}, but with gas both inside and above the funnel cavity (orange and teal regions in Fig. \ref{fig:conegeom}), with $\tau_\mathrm{surr} = \frac{5}{9} \tau_\mathrm{cav}$ }
\label{fig:I_PD_tau_outin}
\end{figure*}

Figure \ref{fig:I_PD_tau_outin} depicts the case with gas simultaneously above and inside the funnel with a total optical depth $\tau_\mathrm{cav}+\tau_\mathrm{surr}$, with both regions having an equal density, $n_\mathrm{cav} = n_\mathrm{surr}$.
Since the inner region spans a larger radius than the outer, the optical depths are different by $\displaystyle \tau_\mathrm{surr} = \frac{5}{9} \tau_\mathrm{cav}$.
The qualitative behavior of the scattered flux is a combination of the previous two cases.
When observing the funnel directly from above, the flux boost from light scattered within the funnel is significant.
Conversely, the escaping flux at higher inclinations mostly consists of light scattered from above the funnel.
Therefore, the polarization from an obscured source is influenced more by scattering from above the funnel than from a medium inside the funnel cavity.

\section{Discussion} \label{sect:discussion}

\subsection{Application to polarimetric data of Cyg X-3}

The PD of the reflected signal in a funnel-type accretion geometry enables constraints to be placed on the funnel opening angle.
Under the assumption of a single-scattering regime, the observed hard-state PD of $20.6\pm0.3\%$ in \mbox{Cyg X-3} translates to the grazing angle $\alpha{\lesssim}15\degr$ \citep[for an observer at $i\approx30\degr$,][]{Veledina2024}.
Our MC simulations show that a similar PD of $\sim$$20\%$ can also be reproduced for larger funnel grazing angles when multiple scatterings are included.
However, significant contributions from multiple scatterings can only be expected at high albedo, which for neutral or partly ionized matter is strongly energy-dependent.
For example, for neutral matter, the absorption opacity depends on the energy roughly as $\sigma_\mathrm{abs}\propto E^{-3}$.
This would result in a pronounced decrease in PD with energy (see Fig. \ref{fig:albedo} for the dependence of PD on albedo), which is not observed.
In addition, if recoil effects become significant, multiple scatterings would further reduce the PD at higher energies.

The observed lack of significant energy dependence in the PD implies that the contribution of multiple scatterings must remain small across the entire IXPE band.
Our simulations show that at $i<40^\circ$ (realistic upper limit for the inclination of \mbox{Cyg X-3}) this requirement is satisfied for $\lambda{\lesssim}0.5$ (Fig.~\ref{fig:albedo}), setting an upper limit for the albedo of the visible surface under our recoil-free approximation.
In this limit, the shape of the reflected spectrum depends on $\lambda(E)$, whereas the PD does not.
This supports the validity of the single-scattering approximation in the hard state.

Drastic changes in the polarization signatures in the intermediate and ultrasoft states (PD $\sim$$10\%$) were previously explained by the presence of an optically thin medium inside or above the funnel cavity \citep{Veledina2024soft}.
Our simulations generally confirm this behavior; however, we found it only necessary to include gas above the funnel directly visible to the distant observer, as the gas inside the cavity had a much weaker influence on the PD (Figs.~\ref{fig:I_PD_tau}, \ref{fig:I_PD_tau_in} and \ref{fig:I_PD_tau_outin}).
In the single-scattering limit, the PD of emission scattered in the medium above the funnel (the scattering component) is set primarily by the observer inclination and is not very sensitive to other parameters.
This suggests that its polarization should be preserved across spectral states.

The observed PD then depends on the relative contribution of this scattering component and the reflection component (the light reflected from the funnel surface).
If the scattered component dominates in both intermediate and ultrasoft states, it could explain the comparable PD in these states.
In our setup, the contribution of the scattered component becomes significant for optical depths of $\tau_\mathrm{surr}\sim0.1$ (Fig.~\ref{fig:PD_tau}).
However, it completely dominates the reflected emission only when the optical depth approaches unity.
In this regime, multiple scatterings become significant, reducing the PD (upper right corner of Fig.~\ref{fig:PD_tau}).
As a result, the PD does not saturate at the observed value of $\sim$$10\%$, which, in contrast, was found to be stable.
This poses a problem for the interpretation of the similar PD between the intermediate and ultrasoft states, as the PD becomes sensitive to variations in the optical depth.
To make matters worse, multiple scatterings would make the PD reduce with energy, while the observed PD is energy-independent.

This problem can be alleviated if the scattered component begins to dominate at lower values of $\tau_\mathrm{surr}$.
In other words, the scattered emission needs to be boosted relative to the reflected emission while keeping the optical depth low.
In the following paragraphs, we discuss a few scenarios that might lead to this.

Since the gas above the funnel is directly exposed to the emission of the primary source, one obvious possibility to make the scattered emission stronger relative to the reflection is to decrease the single-scattering albedo of the funnel surface (see also Fig. \ref{fig:criticaldepth}).
Since we used a value of $\lambda=1/6$ in our simulations, an even lower albedo would be needed.
The low albedo means that the energy is absorbed by the funnel.
The energy might then be partially carried by the outflow or re-emitted by the funnel surface back into the cavity.
If the re-emitted photons remain in the X-ray band, this would add an unpolarized component to the total flux.
However, the upper regions of the funnel surface can have a lower blackbody temperature that produces little X-ray emission, while regions closer to the funnel base can be very hot \citep{Abolmasov2009}.
The upward re-emitted photons would therefore enhance the flux incident on the medium above the funnel, while an observer for whom the source is obscured would only see a weak reflection.

The role of the scattered component may also be enhanced by altering the geometry of the funnel.
Increasing the funnel radius $R$ or decreasing the grazing angle $\alpha$ would reduce the critical optical depth for a given value of the albedo (Fig. \ref{fig:criticaldepth}), as this decreases the reflected flux away from the funnel axis.
However, funnels with larger $R$ require a wider grazing angle to fit the hard-state PD of \mbox{Cyg X-3}, which can lead to increased transverse optical depth and thus multiple scatterings when $R_\mathrm{out}$ is small (see Fig. \ref{fig:PD_tau} and the surrounding discussion).

Yet another possibility to boost the scattered component relative to the reflection is by assuming that the emission is somehow beamed upward from lower in the funnel.
This can be achieved in the aforementioned case of a thermally re-emitting funnel, but also if the lower funnel is perfectly reflecting.
We can compare the angular distribution in our low-albedo simulations with the beamed emission considered in other studies to evaluate its importance.
In our low-albedo model with $\alpha{\sim}10\degr$ and $R=10$, the difference in the total fluxes between $i=0\degr$ and $i=25\degr$ is about three orders of magnitude.
In models that account for geometric beaming within a similar geometry to ours, the difference in bolometric flux between these angles in the free-free re-emission model of \citet{Karmakar2025} is about one order of magnitude, while the perfectly reflecting model of \citet{Dauser2017} exhibits a difference of about two orders of magnitude.
Simulations of the reflection in pulsating ULXs presented in \citet{Mushtukov2021} and \citet{Mushtukov2023} feature a similarly small difference in fluxes across angles, although the model geometry is not directly comparable to ours.
In all of these beaming models, the relative contribution of the scattered component would be diminished.
While ignoring the energy dependence of the radiation, this would lead to a critical optical depth of well above unity.
This reiterates our conclusion that at least the upper parts of the funnel surface must be strongly absorbing for the scattered component to be dominant.

\subsection{Physical scenario for the state changes in Cyg X-3}

\begin{figure}
\begin{center}
\includegraphics[width=\linewidth]{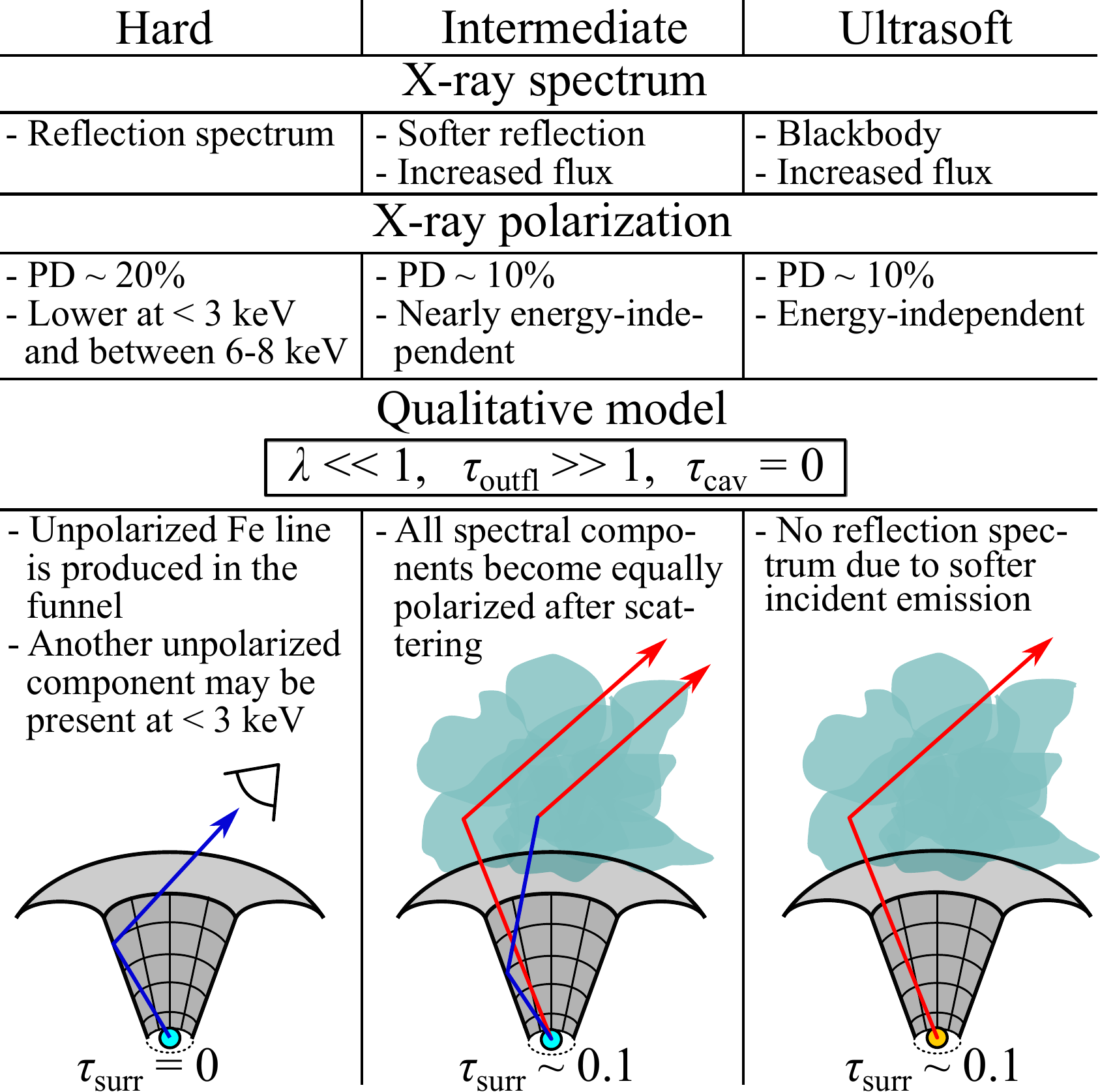}
\end{center}
\caption{Illustrative diagram of the qualitative model for the state changes in \mbox{Cyg X-3}. The observed features in the hard, intermediate, and ultrasoft states are listed above, with the corresponding model sketched below. The sketches highlight the reflection from the funnel surface (blue), the scattering from the medium above the funnel (red), and the hardness of the intrinsic spectrum (teal, orange).}
\label{fig:illustration}
\end{figure}

The strikingly similar polarization in the intermediate and ultrasoft states of \mbox{Cyg X-3} implies a clear continuity between the two states.
Given that both states feature a PD consistent with single-scattering by an optically thin medium above the funnel, it seems that the presence of the scattering medium may be key in explaining the spectral transitions.
Therefore, we propose that both states feature the same scattering medium.
The addition of a scattered component would be consistent with the increased X-ray flux observed during both states.
Furthermore, the medium can explain the energy-independent PD observed during both states; during the hard state, the PD is reduced near 6 keV due to dilution by the iron fluorescent line \citep{Veledina2024}, but the line emission would have the same polarization as the continuum if everything is scattered by the same medium. 
Figure \ref{fig:illustration} summarizes this scenario for the spectral states.

As noted by \citet{Veledina2024soft}, the softening of the spectrum  in the hard-to-intermediate state transition in \mbox{Cyg X-3} can be attributed to the medium above the funnel redirecting some of the concealed soft emission toward the observer.
Unlike the partially ionized funnel surface, the completely ionized medium would not alter the spectral shape of the initial emission.
Therefore, the spectrum of the scattered component would be similar to the likely softer spectrum of the central emission.
Even if the intrinsic spectrum does not change, the addition of a softer scattered component may thus be able to explain the spectral transition as seen by the observer.
Since the reflection spectrum nevertheless remains prominent in the intermediate state, the obscured regions of the funnel surface may feature significant reflection.

If the spectrum in both the intermediate and ultrasoft states is formed by scattering the obscured emission toward the observer, it would indicate that the ultrasoft state has a softer intrinsic spectrum.
This would be consistent with the lack of a reflection component in the ultrasoft spectrum, as a supply of hard X-rays is required for down-scattering to be relevant.
The softening of the spectrum and the quenching of the jet during the intermediate-to-ultrasoft state transition are reminiscent of the intermediate-to-soft state transition in regular black hole XRBs, which is associated with an increase in the accretion rate \citep{Fender2004}.

An increased mass-loss rate of the WR wind would naturally explain the increased optical depth of gas above the funnel cavity.
Any alterations to the geometry caused by the wind-outflow interaction would be observable as a systematic shift in the PA.
The difference in PA between the IXPE observations is not statistically significant, although the orbital variability of the PA does appear to change in the ultrasoft state \citep{Veledina2024soft}.

The duration of the intermediate-to-ultrasoft state transition, which can last up to dozens of days \citep{Szostek2008}, may be consistent with the timescale needed for the increased mass supply to reach the inner disk.
The accretion timescale for a standard geometrically thin disk at radius $r$ \citep{Lightman74} is
\begin{equation}
    t_\mathrm{acc} = \frac{r^{3/2}}{\alpha_\mathrm{v}\sqrt{GM}} \left( \frac{z_0}{r} \right)^{-2},
\end{equation}
where $\alpha_\mathrm{v}$ is the effective viscosity, $M$ is the compact object mass, and $z_0$ is the half-height of the accretion disk at $r$.
Assuming values of $M\sim3~\mathrm{M}_\odot$ and $r\sim3\times10^{10}~\mathrm{cm}$ \citep[about one-tenth of the orbital separation; ][]{Zdziarski2012} with $z_0/r \sim 0.1$, the timescale can be comparable to the length of the transition if $\alpha_\mathrm{v}$ is on the order of $10^{-2}$--$10^{-3}$.
The radiative general relativistic magnetohydrodynamical (RGRMHD) simulations of super-Eddington accretion in \citet{Fragile2025} featured such low values for the $\alpha_\mathrm{v}$ parameter, yet it is unclear how applicable the standard estimate of the timescale is for a wind-fed supercritical disk.
The timescale of the transition may otherwise be related to the duration of the enhanced mass-loss rate or the interaction between the WR wind and the disk outflow.

Since the scattering medium above the funnel should no longer be present when the system returns to the hard state, we expect that the medium is ejected during the major radio flares that follow the ultrasoft state.
Therefore, we predict that the PD during the flaring state may initially increase as a result of relativistic aberration if the scattering medium has a large outflowing velocity.
Later during the outburst, the PD should settle toward its hard-state value.
We emphasize that this is a speculative prediction, and that testing this scenario for the spectral transitions would require energy-dependent simulations.

\subsection{Other sources}

While \mbox{Cyg X-3} is hitherto the only XRB observed to have a strong X-ray polarization consistent with an obscuring outflow, we anticipate that similar objects may be observed with IXPE and, in the future, with the enhanced X-ray Timing and Polarization mission \citep[eXTP,][]{Zhou2025extp, Zhang2025}.
Other known Galactic XRBs that may be obscured have higher inclinations, so we expect their PD to be high as long as the reflected radiation is not diluted by emission leaked through the funnel.
For example, the inclinations of \mbox{SS 433} \citep{Fabrika04},
\mbox{V404 Cygni} \citep{Khargharia2010}, \mbox{GRS 1915+105} \citep{Reid2023} and \mbox{V4641 Sagittarii} \citep{MacDonald2014} have been measured at roughly $78\degr$, $67\degr$, $64\degr$, and $72\degr$, respectively.
If the funnel surface is optically thick to absorption or if the light is scattered from an optically thin medium above the funnel, we expect the PD of the reflected emission to be $>50\%$ at these inclinations.
In the case of SS~433, regular precession of the accretion disk with the amplitude of $21\degr$ should result in noticeable modulation of the reflected PD with the precessional phase. 

Several high-inclination systems show clear signatures of obscuration and outflows, yet their observed PDs remain lower than expected.
For instance, the high-inclination black hole XRB \mbox{4U 1630$-$47} was found to have an energy-dependent PD, increasing from $5\%$ to $10\%$ in the IXPE band, while remaining stable across different spectral states (soft and steep power-law state). 
This was explained by scattering from a geometrically thick accretion disk or a disk outflow \citep{Ratheesh2024,RodriguezCavero2023}.
Another example is the neutron star XRB \mbox{GX 13+1}, which was recently revealed to be obscured and intrinsically bright \citep[accreting at near- or slightly above the Eddington limit,][]{XRISMgx13}.
However, the simultaneous IXPE observations found a total PD of only about $2.5\%$ \citep{Bobrikova24new} despite of the high system inclination ${\sim}60\degr$--$80\degr$ \citep{DiazTrigo2012}.
The properties of the obscuring envelope were found to be consistent with thermal-radiative winds with an electron scattering optical depth near unity.

In summary, both \mbox{4U 1630$-$47} and \mbox{GX 13+1} appear to have high accretion rates and are expected to be surrounded by a scattering medium, yet their X-ray emission is much less polarized than in \mbox{Cyg X-3}.
This can be attributed to an accretion rate that is only nearly Eddington or mildly super-Eddington, which results in a less optically thick outflow with a larger opening angle, leading to greater dilution of the highly polarized reflected flux by the (partly absorbed) unpolarized incident emission.

In the highly and stably super-Eddington source \mbox{SS 433} \citep[reviewed in][]{Fabrika04}, the X-ray emission is dominated by direct emission from its unique baryonic jets \citep{1986MNRAS.222..261W, 1996PASJ...48..619K,Marshall2002,2016MNRAS.455.1414K,2019AstL...45..299M}, although signatures of reflected emission have also been observed and modeled \citep{2010MNRAS.402..479M,Medvedev2018,Middleton2021}.
Because of the jets emission, the polarization of the observed emission should be strongly diluted.
The exact geometry of the various components that contribute to the continuum \mbox{SS 433} is debated \citep[e.g.,][ for discussion of various geometries]{2010MNRAS.402..479M,Middleton2021,Medvedev2018}, making it difficult to robustly predict the PD and PA and their temporal variations.
Broadly speaking, if the observed PD will positively correlate with the relative contribution of the reflected continuum to the spectrum, one can suggest a picture similar to the one considered here, where the reflected PD is high and changes gently, while the reflected flux, $F_{I,\mathrm{refl}}/F_{0}$, is small ($10^{-5}-10^{-4}$) but changes significantly with changing inclination.
The lack of such correlations or inconsistent values of PA or PD will be indicators of more complex scattering geometry, i.e., as a result of scattering in an extended wind or re-processing of the jets' base emission.
A combination of quasi-simultaneous polarimetric and spectroscopic X-ray observations of \mbox{SS 433} at different precessional phases will be key for disentangling contributions of the various emission components.   

\subsection{Mechanisms for geometric beaming}

Throughout our discussion, we have assumed a single-scattering albedo of much less than unity to explain the high PD of \mbox{Cyg X-3} in its hard state.
We find that geometric beaming does not occur via reflection unless the albedo is very close to unity, and thus favors re-emission of the absorbed radiation as the mechanism behind collimation.
The assumed presence of absorption nevertheless has caveats, as the funnel surface is expected to be very hot (${\sim}10^7~\mathrm{K}$) and subject to intense irradiation.
In a highly ionized gas, free-free absorption and Compton scattering dominate the opacity.
The classical ratio between the free-free and Thomson cross sections is \citep{Rybicki1986}
\begin{equation}
    \frac{\sigma_\mathrm{ff}}{\sigma_\mathrm{Th}} = 1.24\times10^{-23}~ n_\mathrm{e}Z^2  T_7^{-1/2}  E_\mathrm{keV}^{-3} \left[ 1 - \exp\left(\frac{-1.16E_\mathrm{keV}}{T_7}\right) \right],
\end{equation}
where $n_\mathrm{e}$ is the electron number density in $\mathrm{cm}^{-3}$, $Z$ is the atomic number, $E_\mathrm{keV}$ is the photon energy in kiloelectronvolts, and $T_7$ is the temperature in units of $10^7~\mathrm{K}$.
In order for the single-scattering albedo to be much less than unity in the IXPE range, the visible portion of the funnel surface would have to be unrealistically dense \citep[$n_\mathrm{e}\gtrsim10^{23}~\mathrm{cm^{-3}}$, while in recent simulations the highest outflow density was about ${\sim}10^{20}~\mathrm{cm^{-3}}$;][]{Zhang2025b} or have temperatures below $\lesssim10^6~\mathrm{K}$.
It is therefore likely that parts of the funnel are purely reflecting.
A lower albedo may be possible in the upper section of the funnel surface if it is less ionized.
However, photoelectric absorption is difficult to quantify without self-consistent modeling.

The high PD observed in \mbox{Cyg X-3} implies that the upper region of the funnel surface is only partly ionized and relatively cool (${\sim}10^5 ~\mathrm{K}$), which can be consistent with the analytical estimate of $T\propto r^{-1/2}$ \citep{Abolmasov2009} if the outflow extends thousands of gravitational radii away from the central source.
However, some RGRMHD simulations seem to exhibit a slower decline of the temperature with distance \citep{Kitaki2021, Zhang2025b}.
Spectra of some ULXs contain wide absorption features with mildly relativistic outflow velocities, implying that a relatively low ionization is possible far away from the compact object \citep{Middleton2014}.
Thermal re-emission from this neutral matter would additionally dilute the polarized reflection at low energies, which may be consistent with the lower PD at energies below 3 keV observed during the hard state of \mbox{Cyg X-3}  \citep{Veledina2024, Mikusincova2025}.

We conclude that a more detailed model is needed to comprehensively explain geometric beaming in ULXs.
While in this work we only considered the simple case of a constant albedo across the entire outflow, the lower parts of the funnel surface are likely reflection dominated, whereas absorption and re-emission may become important higher in the funnel.
Additionally, light transmitted through the outer boundary of the outflow can significantly reduce the beaming factor from the ideal geometric estimate.
Furthermore, a fraction of the light exiting the cavity escapes toward the sides.
Our model geometry allowed for significant leakage through the outflow, so the flux boost in our $\lambda=1.0$ simulations was significantly smaller than that of the fully opaque funnel reflection modeled in \citet{Dauser2017}.
Relativistic outflow velocities can increase the amount of collimation, although it may also enhance leakage through the upper boundary of the outflow.
This mechanism faces difficulties in \mbox{Cyg X-3} and \mbox{GX 13+1}, since their spectra do not feature relativistic velocities \citep{XRISMgx13}.
We leave further discussion of the beaming to a future work.

\section{Summary} \label{sect:summary}

In this work, we explored how different outflow properties influence the flux and polarization of either black holes or neutron stars accreting in the super-Eddington regime, using \mbox{Cyg X-3} as a case study.
We performed monochromatic MC simulations of radiative transfer in the funnel.
We also accounted for a low-density scattering medium within and above the funnel cavity.

We found that the emission reflected by the funnel surface is strongly polarized only when absorption suppresses multiple scatterings; as a result, most of the flux is absorbed.
Conversely, the funnel reflection can only collimate the flux when little absorption is present.
This shows that obscured ULXs cannot be strongly polarized if they are beamed purely by a reflecting funnel surface.
The high PD observed in \mbox{Cyg X-3} thus indicates that the visible portion of the funnel is only partly ionized. 

We found that the presence of the optically thin medium above the funnel can qualitatively explain all the spectral states in \mbox{Cyg X-3} observed by IXPE.
We considered a narrow ($\alpha < 20^\circ$) and moderately extended ($R<100$) funnel with a large scattering optical depth ($\tau_\mathrm{outfl}=18$) and a single-scattering albedo corresponding to neutral matter ($\lambda=1/6$).
We tested different optical depths of gas both inside and above the funnel, but found the medium inside the cavity far less consequential.
If the optical depth of the medium above the funnel is vanishingly small ($\tau_\mathrm{surr}\lesssim10^{-2}$) during the hard state but becomes substantial during the intermediate and ultrasoft states ($\tau_\mathrm{surr}\sim0.1$), it would account for the increased flux observed during the two latter states, as the medium scatters some of the otherwise obscured emission toward the observer.

Unlike the partially ionized funnel surface that produces the hard reflected spectrum, this fully ionized optically thin medium scatters emission without modifying its spectrum (apart from the recoil effect).
The hard and intermediate states may therefore have similar intrinsic spectra, but the intermediate state would appear softer to a distant observer due to the light scattered by the medium above the funnel.
In contrast, the ultrasoft state cannot be explained unless the source has a softer intrinsic spectrum.

The medium above the funnel can also explain the reduced PD observed during the intermediate and ultrasoft states, since it leads to lower scattering angles than the funnel reflection observed during the hard state. 
To match the observed PD of ${\sim}10$\% in these states, the optical depth of the scattering medium may have to be rather large ($\tau_\mathrm{surr}\sim0.5$) depending on the angular distribution of the flux.

We found that geometric beaming produced by reflection from the funnel surface was relatively weak in our simulations, even for a purely reflecting funnel. 
This was caused by two effects.
Firstly, a large fraction of photons escape through the outer boundary of the outflow rather than the funnel cavity.
Secondly, some of the light exits the cavity at high angles with respect to the funnel axis.
While the effect is undoubtedly influenced by our model geometry, it demonstrates that beaming in ULXs may be appreciably less efficient than the limiting case in which all photons are collimated along the funnel. 

\begin{acknowledgements}
We thank Juri Poutanen and the anonymous referee for the helpful comments. 
VA and AV were supported by the Research Council of Finland grants 355672, 372881 and the Centre of Excellence in Neutron-Star Physics (grant 374064).
VA acknowledges support from the Turku University Foundation.
Nordita is supported in part by NordForsk. 
IK was supported by the Simons Foundation via the Simons Investigator Award to A. A. Schekochihin.
\end{acknowledgements}

\bibliographystyle{aa}
\bibliography{references}

\begin{appendix}

    \section{Surface integral for the reflected flux} \label{sect:reflection}

\begin{figure}
\begin{center}
\includegraphics[width=0.7\linewidth]{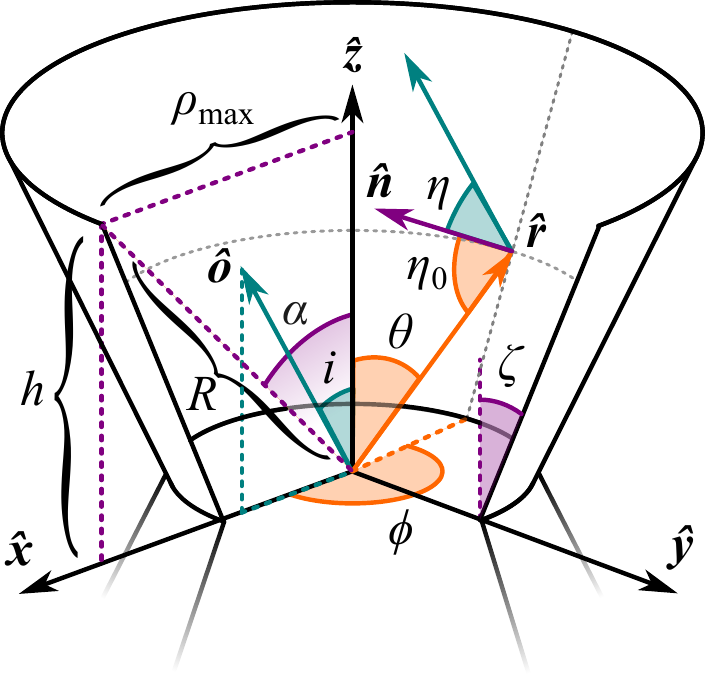}
\end{center}
\caption{Illustration of the reflection geometry.}
\label{fig:scattering_geom}
\end{figure}

    To calculate the reflection from the surface of the funnel \citep[see][]{Veledina2024}, we employ a coordinate system where the $z$-axis is aligned with the symmetry axis of the funnel, $\Vec{\hat \Omega} = (0,0,1)$, and the observer lies on the $x$-$z$ plane.
    We define the direction toward the observer as $\Vec{\hat o} = (\sin i, 0, \cos i)$.
    The light is emitted from the origin toward the funnel surface along the vector $\Vec{\hat r} = (\sin \theta \cos \phi, \sin \theta \sin \phi, \cos \theta)$, where $\theta$ and $\phi$ are the colatitude and azimuth of the reflecting point, and then scatters toward the observer direction.
     The cosine of the scattering angle is thus
    \begin{equation}
        \mu = \Vec{\hat r} \cdot \Vec{\hat o} = \cos i \cos \theta + \sin i \sin \theta \cos \phi.
    \end{equation}
    The geometry is shown in Fig. \ref{fig:scattering_geom}.

    The funnel geometry is most conveniently expressed in cylindrical coordinates.
    In the $z$-direction, the funnel reaches up to a height of $h=R\cos\alpha$ where its cylindrical radius in the $x$-$y$ plane is $\rho_\mathrm{max}=R\sin \alpha$.
    The cylindrical radius of the funnel surface as a function of $z$ in units of the funnel base radius is
    \begin{equation}
        \rho = |z| \tan \zeta + 1,
    \end{equation}
    where $\tan \zeta = (\rho_\mathrm{max}-1)/h$ is the tangent of the cone half-opening angle.
    At a given point on the surface, the surface normal of the funnel is $\Vec{\hat n} = (-\cos \zeta \cos \phi, -\cos \zeta \sin \phi, \mathrm{sgn}(z) \sin \zeta)$.
    The $z$-component of the normal depends on the sign of the $z$-coordinate of the considered point, since the funnel geometry is mirrored around the \mbox{$x$-$y$} plane.
    The cylindrical coordinates $z$ and $\rho$ can be converted to the colatitude using the relations $\sin \theta = \rho/\sqrt{\rho^2 + z^2}$ and $\cos \theta = z/\sqrt{\rho^2 + z^2}$.
    
    The cosine of the angle between the incident light and the surface normal is
    \begin{equation}
        \eta_0 = -\Vec{\hat r} \cdot \Vec{\hat n} =-\sin (\zeta - \mathrm{sgn}(z) \theta),
    \end{equation}
    and the cosine of the angle between the normal and the observer direction is
    \begin{equation}
        \eta = \Vec{\hat o} \cdot \Vec{\hat n} = \mathrm{sgn}(z)\cos i \sin \zeta - \sin i \cos \zeta \cos \phi.
    \end{equation}
    The polarization basis is formed by the projection of $\Vec{\hat \Omega}$ on the sky,
   \begin{align}
       \Vec{\hat e}_1 &= \frac{\Vec{\hat \Omega} - \Vec{\hat{o}}\cos i  }{\sin i} = (-\cos i, 0, \sin i), \\
       \Vec{\hat e}_2 &= \frac{ \Vec{\hat{o}} \times  \Vec{\hat \Omega}}{\sin i} = (0, -1, 0),
   \end{align}
   and the polarization pseudo-vector is parallel to the normal of the scattering plane:
   \begin{equation}
       \Vec{\hat p} = \frac{\Vec{\hat o} \times \Vec{\hat r}}{|\Vec{\hat o} \times \Vec{\hat r}|}.
   \end{equation}
   The PA of light scattered from point $\Vec{\hat r}$ is thus computed as
   \begin{align}
       \cos \chi &= \Vec{\hat e_1} \cdot \Vec{\hat p} = \frac{\sin \theta \sin \phi}{\sqrt{1-\mu^2}}, \\
       \sin \chi &=\Vec{\hat e_2} \cdot \Vec{\hat p} = \frac{\sin i \cos \theta - \cos i \sin \theta \cos \phi}{\sqrt{1-\mu^2}}.
   \end{align}

Finally, the Stokes flux reflected from the surface of the funnel is \citep[][p. 146]{Chandraskhar1960}
\begin{equation} 
    \left( \begin{matrix}
        F_{I, \mathrm{refl}} \\
        F_{Q, \mathrm{refl}}
    \end{matrix} \right) = 
    F_0\frac{3\lambda}{16\pi}
    \int\limits_{-h}^{+h}  \int\limits_{0}^{2\pi} \frac{\rho}{z^2 + \rho^2}
 \left( \begin{matrix} 
      1+\mu^2\\
      (1-\mu^2) \cos 2 \chi
    \end{matrix} \right)
    \frac{\eta \eta_0}{\eta + \eta_0}
    \, \frac{\mathrm{d}\phi\,\mathrm{d}z}{\cos \zeta}.
\end{equation}
Although the integral as written covers the full surface of the funnel, it is subject to three visibility conditions:
(i) the reflected light cannot propagate into the funnel surface, (ii) the line of sight must not intersect the upper rim of the funnel, and (iii) light scattered from below $z<0$ must pass through the central hole in the funnel cavity.
Condition (i) is simply satisfied by
\begin{equation} \label{eq:conditioni}
    \eta >0,
\end{equation}
since $\eta_0$ is always positive.
For condition (ii), the light trajectory at $z=h$ must be within the funnel cavity, i.e., at a cylindrical radius less than $\rho_\mathrm{max}$:
\begin{equation}\label{eq:conditonii}
    \rho^2 + (h-z)^2 \tan^2 i + 2\rho (h-z) \tan i \cos \phi < \rho_\mathrm{max}^2.
\end{equation}
Condition (iii) has a similar expression, as the trajectory at $z=0$ must pass within the inner radius of the frustum:
\begin{equation}\label{eq:conditoniii}
    \rho^2 + z^2 \tan^2 i - 2z\rho \tan i \cos \phi < 1.
\end{equation}
To follow these conditions, we found the visible range of $\phi$ as a function of $z$ by inverting Eqs. (\ref{eq:conditioni}--\ref{eq:conditoniii}), and calculated the overlap of the three conditions.
We then integrated $z$ from the lowest visible point up to $h$.

The lowest visible point along the $z$-axis, $z_\mathrm{min}$, is always located at the azimuth of $\phi=\pi$ opposite to the observer.
From condition (i), we find $\sin(i +\mathrm{sgn}(z)\zeta)>0$.
This is always true when $z>0$, but the funnel is not visible below $z<0$ if $i<\zeta$.
According to condition (ii), the lowest visible point is $z_\mathrm{min}\leq0$ when $i \leq \arctan[(1+\rho_\mathrm{max})/h]$.
Combining these, we find that the lower half of the funnel is only visible between $\zeta<i<\arctan[(1+\rho_\mathrm{max})/h]$, and that $z_\mathrm{min} = 0$ when $i<\zeta$.
Condition (iii) has no impact on $z_\mathrm{min}$ in this inclination range, since it is automatically satisfied when $z_\mathrm{min}\geq2/(\tan \zeta - \tan i)\geq-h$.
Outside of $i<\zeta$, the visibility is thus entirely determined by condition (ii):
\begin{equation}
    z_\mathrm{min} = \frac{h\tan i - \rho_\mathrm{max} - 1}{\pm \tan \zeta + \tan i},
\end{equation}
where the positive sign of $\tan\zeta$ corresponds to inclinations where the lower half of the funnel is concealed ($i>\arctan[(1+\rho_\mathrm{max})/h]$), and the negative sign to inclinations where the lower half is visible (vice versa).

\end{appendix}

\end{document}